%% file: datacentergym.tex
\documentclass[conference]{IEEEtran}

\IEEEoverridecommandlockouts

\usepackage{multirow}
\usepackage{cite}
\usepackage{amsmath,amssymb,amsfonts}
\usepackage{graphicx}
\usepackage{textcomp}
\usepackage{xcolor}

\usepackage{tikz}
\usetikzlibrary{arrows.meta,positioning,calc,fit}

\usepackage{algorithm}
\usepackage{algcompatible} 

\usepackage[caption=false,font=footnotesize]{subfig}

\usepackage{enumitem}
\setlist[itemize]{leftmargin=*,noitemsep,topsep=2pt}
\setlist[enumerate]{leftmargin=*,noitemsep,topsep=2pt}
\usepackage{balance} 

\def\BibTeX{{\rm B\kern-.05em{\sc i\kern-.025em b}\kern-.08em T\kern-.1667em\lower.7ex\hbox{E}\kern-.125emX}}



\begin{document}

\title{DataCenterGym: A Physics-Grounded Simulator for Multi-Objective Data Center Scheduling}

\author{
\IEEEauthorblockN{Nilavra Pathak\IEEEauthorrefmark{1}, Samadrita Biswas\IEEEauthorrefmark{2}, and Nirmalya Roy\IEEEauthorrefmark{3}}

\IEEEauthorblockA{\IEEEauthorrefmark{1}Marketing Data Science, Expedia Group}
\IEEEauthorblockA{\IEEEauthorrefmark{2}Analytical Development, Curia Global}
\IEEEauthorblockA{\IEEEauthorrefmark{3}Department of Information Systems, University of Maryland, Baltimore County}

\IEEEauthorblockA{\IEEEauthorrefmark{1}npathak@expediagroup.com,
\IEEEauthorrefmark{2}samadrita.biswas@curiaglobal.com,
\IEEEauthorrefmark{3}nroy@umbc.edu}
}

\maketitle

\begin{abstract}
Modern datacenters schedule heterogeneous workloads across geo-distributed
sites with diverse compute capacities, electricity prices, and thermal
conditions. Compute utilization, heat generation, cooling demand, and energy
consumption are tightly coupled, yet most existing schedulers abstract these
effects and treat them independently.

We present \textit{DataCenterGym}, a physics-grounded simulation environment for
job scheduling in geo-distributed data centers, designed as a reusable testbed
for future research. The simulator integrates compute queueing, building
thermal dynamics, localized HVAC behavior, and temperature-dependent service
degradation within a Gymnasium-compatible interface. We also develop a
Hierarchical Model Predictive Control (H-MPC) scheduling algorithm that
performs distributed job placement while explicitly accounting for thermal and
power dynamics. Through experiments on nominal operation and workload
sensitivity, we demonstrate how H-MPC improves scheduling
performance relative to baseline schedulers.
\end{abstract}


\input{Introduction}
\input{Related}
\input{Environment}
\input{Allocation}

\input{Experimental_Setup}

\input{Results}
\input{Conclusion}


\balance
\bibliographystyle{IEEEtran}
\bibliography{bib}

\end{document}

%% file: Introduction.tex
\section{Introduction}

The rapid expansion of hyperscale and edge data centers is driven by growing demand for large-scale AI training, cloud services, and latency-sensitive applications. These facilities account for a substantial and rising share of global electricity consumption, posing significant challenges in thermal management, energy provisioning, and sustainable operations \cite{maji2025data, kamiya2025data}. Operators must continuously adapt to fluctuating workloads and heterogeneous infrastructure while maintaining thermal safety, hardware reliability, and performance across geographically distributed sites.

Traditional resource managers were designed to optimize utilization and latency, but largely ignore the environmental interactions. Job execution generates heat proportional to resource consumption, raising internal temperature and driving cooling demand. Due to thermal inertia, these effects unfold over minutes to hours rather than instantaneously~\cite{moore2005making}. Hence, schedulers optimizing for short-term throughput can inadvertently cause thermal violations. This delayed effects can harm the hardware devices over time.

Geographic distribution further amplifies these challenges. A GPU workload executed in Phoenix during summer requires substantially more cooling than the same workload in Seattle. Most production schedulers model data centers as interchangeable capacity pools, without explicitly accounting for thermal and power dynamics \cite{verma2015large,hindman2011mesos}. High-density accelerators further exacerbate the problem: GPU clusters generate approximately $3$--$4\times$ higher heat density per node than CPU clusters~\cite{nvidia_datacenter_power,  zhang2020datacenterthermal}, placing increased stress on cooling systems and increasing the risk of thermal throttling under sustained load. This simultaneously increases the energy usage, and hence the carbon footprint and cost of operations.   

 \textbf{The Challenge:} Prior work has explored power-aware scheduling~\cite{verma2009pmapper} and control-theoretic approaches to thermal management in data centers~\cite{lazic2018data}. However, there is no comprehensive dataset or evaluation framework that jointly captures the full range of environmental and operational interactions in modern data centers. As a result, simulation-based platforms have emerged as a practical means to evaluate control-theoretic and reinforcement learning approaches for scheduling~\cite{li2023sustaingym}. Existing simulators, however, do not support closed-loop physical dynamics. Consequently, the interactions among scheduling decisions, temperature evolution, cooling demand, and thermal throttling under temperature spikes cannot be fully explored. Evaluating policies that explicitly reason about these inter-dependencies requires a unified simulation environment that integrates workload dynamics, thermal physics, and cooling control.
 This setting introduces several fundamental challenges:

\begin{itemize}
    \item \textbf{Coupled objectives:} Scheduling decisions must balance throughput, latency, energy efficiency, and thermal safety under shared physical constraints, where improving one objective can degrade others.

    \item \textbf{Delayed physical effects:} Thermal inertia and cooling response times induce delayed thermal stress. The impact of a scheduling decision may emerge gradually over time.

    \item \textbf{Heterogeneity and scale:} Geo-distributed data centers exhibit diverse climate conditions, hardware characteristics, power limits, and electricity pricing, resulting in location-dependent thermal and cost profiles for identical workloads.

    \item \textbf{Realistic workload modeling:} Synthetic or simplified job traces fail to capture key characteristics of real workloads. They miss hardware affinity constraints and the temporal arrival patterns observed in production systems.

\end{itemize} 

\textbf{Our Contributions.} This paper presents \emph{DataCenterGym}, a physics-aware simulation environment for geo-distributed data centers. It enables the evaluation of scheduling policies under coupled thermal, power, and workload dynamics. Our contributions are as follows:

\begin{enumerate}
    \item Simulator modeling heat generation, thermal evolution, localized HVAC control, and thermal throttling across heterogeneous compute clusters in geographically distributed data centers.

    \item Integration of production-scale workload traces capturing realistic job arrivals, execution durations, resource demands, and hardware affinity constraints.

    \item A safety-constrained Model Predictive Control algorithm that jointly schedules jobs and thermostat setpoints to achieve optimal performance under physical constraints.

    \item Experimental evaluations across nominal operation and varying workload intensities. These experiments reveal how physical couplings materially influence scheduling policy performance.

\end{enumerate}

\textit{DataCenterGym} enables principled evaluation of datacenter scheduling algorithms under realistic thermal and power constraints without requiring access to production infrastructure.

%% file: Related.tex
\section{Related Works}

\textbf{Geo-distributed scheduling:}
Geographic load balancing reduces operational cost by routing workloads across
regions with heterogeneous electricity prices \cite{qureshi2009cutting}, with
extensions for deadlines, data locality, and capacity constraints
\cite{wu2025survey,yuan2020profit,niu2020gmta}. Most approaches treat power and thermal limits as fixed feasibility constraints
rather than dynamic states coupled to workload decisions.

\textbf{Datacenter resource managers:}
Production systems such as Borg \cite{verma2015large} and Mesos
\cite{hindman2011mesos} prioritize scalability, fairness, and utilization, while
thermal management is handled reactively through throttling or emergency cooling
\cite{moore2005making}. Consequently, schedulers are rarely evaluated for temporal thermal or energy
effects.

\textbf{Thermal- and power-aware scheduling:}
Temperature-aware placement mitigates hotspots and cooling demand
\cite{moore2005making}, and joint workload--cooling coordination reduces energy
use \cite{nejad2020joint}. Power-aware schedulers such as PMapper
\cite{verma2009pmapper} enforce instantaneous limits but do not account for
thermal inertia. Anticipatory control
\cite{lazic2018data,banerjee2010cooling}, together with data-driven forecasting of
building-level thermal dynamics \cite{pathak2018forecasting, pathak2019bayesian},
motivates predictive models used in control-oriented scheduling.

\textbf{Carbon-aware scheduling:}
Schedulers exploit spatial and temporal variation in grid carbon intensity to
reduce emissions \cite{souza2023casper,hanafy2024gaia,zhai2025f2swss,chen2025carbon},
but rely on coarse-grained power models that omit cooling and thermal constraints.
Hierarchical demand-response mechanisms coordinate operators and tenants under
power limits \cite{xu2017hierarchial}.

\textbf{Learning-based scheduling:}
Reinforcement learning has been applied to cluster-scale scheduling
\cite{mao2016deeprm} and extended to geo-distributed systems
\cite{niu2020gmta}. Policy performance depends critically on simulator fidelity;
omitting thermal inertia, cooling response, or power limits risks poor
generalization.

\textbf{Simulation platforms:}
CloudSim-style simulators support geo-distribution and energy modeling
\cite{calheiros2011cloudsim,kliazovich2012greencloud,alves2021gdsim} but decouple
scheduling from delayed physical feedback. SustainGym
\cite{li2023sustaingym} and Clockwork~\cite{valdez2021clockwork} benchmark
sustainability-aware optimization while abstracting job placement and thermal
throttling. OpenDC \cite{klint2020opendc} emphasizes scalability but omits
closed-loop thermal and cooling interactions.

\textbf{Positioning:}
DataCenterGym provides a closed-loop, cyber--physical simulation of
geo-distributed data centers that jointly models job scheduling, queueing,
thermal dynamics, cooling control, and power constraints, enabling evaluation
beyond capacity-only abstractions.

%% file: Environment.tex
\section{DataCenterGym Environment}
\label{sec:environment}

\textit{DataCenterGym} models online job allocation as a controlled, discrete-time
stochastic environment whose state transitions are driven by workload arrivals,
job execution, thermal dynamics, cooling control, and power evolution. By varying
workload characteristics, ambient conditions, and physical parameters, the
environment spans diverse operating regimes while preserving a consistent
decision structure. This section defines the state, action space, and system
dynamics.

\begin{figure}[H]
\captionsetup{font=scriptsize}
\centering
\includegraphics[width=\linewidth]{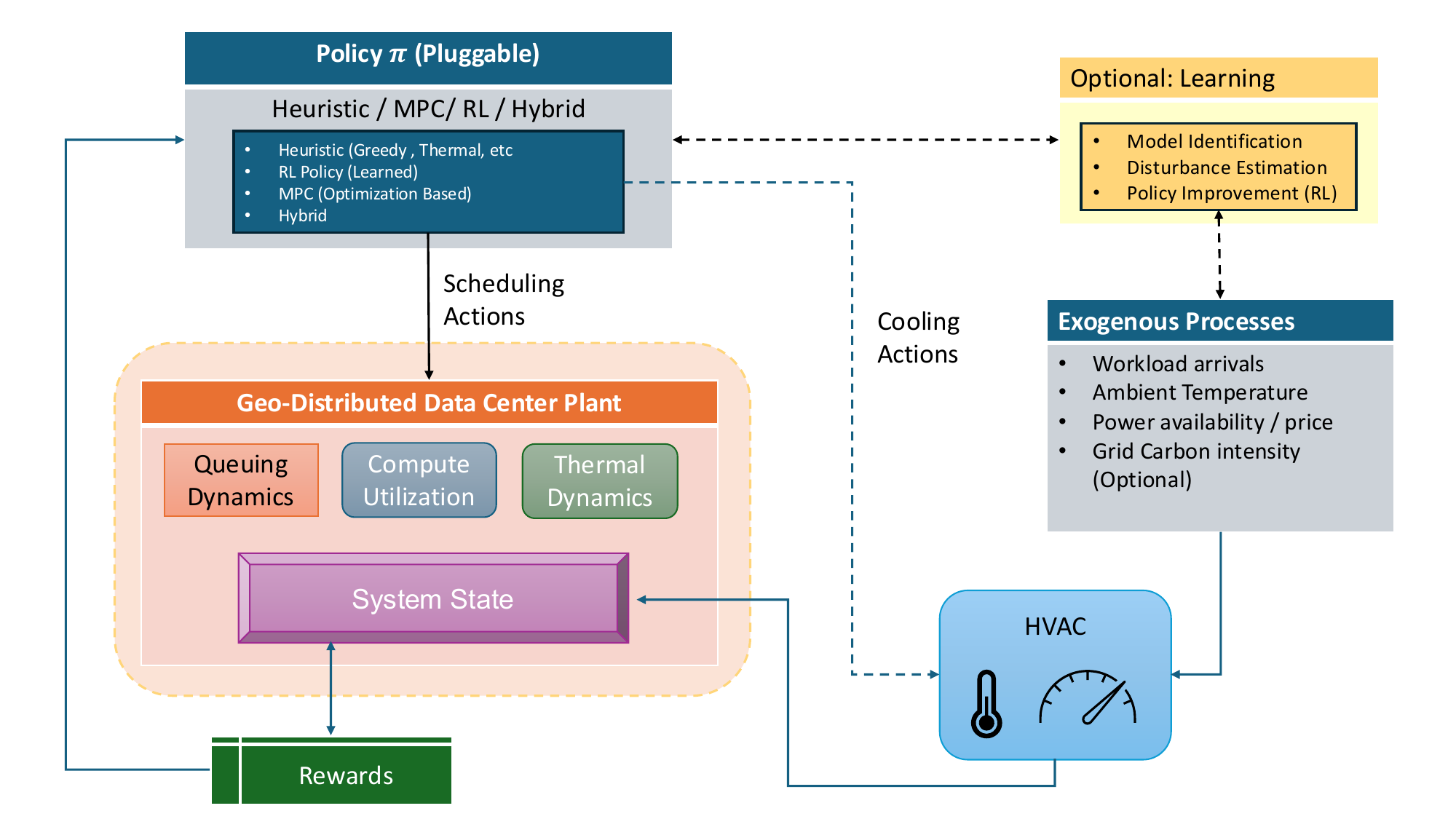}
\caption{Closed-loop interaction in \textit{DataCenterGym}. The scheduler observes
the system state, selects job assignment and cooling actions, and the environment
advances via coupled workload execution, thermal dynamics, and power evolution.}
\label{fig:dcgym_arch}
\end{figure}

\subsection{Problem Formulation}

We consider an online scheduling problem over $C$ compute clusters distributed
across $D$ geo-distributed datacenters. Time is discretized into intervals
$t = 0, \dots, T$ with step size $\Delta t$. Jobs arrive over time with
heterogeneous resource demands, execution durations, priorities, and hardware
affinities, and must be scheduled subject to compute capacity, thermal, and power
constraints.

\subsubsection{State Space}

The system state comprises cluster-level scheduling variables and
datacenter-level environmental variables.
At the cluster level, each cluster $i \in \{1,\dots,C\}$ is characterized by:
(i) available electrical power capacity $p_{i,t}$,
(ii) effective compute capacity $c_{i,t}$, and
(iii) queue length $q_{i,t}$. Cluster utilization $u_{i,t}$ is defined as the total active resource demand
assigned to cluster $i$ (~Section~\ref{sec:job_execution}) and is therefore not
included as an independent state variable.
At the datacenter level, each datacenter $d \in \{1,\dots,D\}$ maintains:
(i) an internal thermal state $\theta_{d,t}$, representing a control-level
temperature proxy,
(ii) ambient (outdoor) temperature $\theta^{\mathrm{amb}}_{d,t}$, and
(iii) electricity price $\psi_{d,t}$.
These variables are shared across all clusters hosted within the same
datacenter.

The scheduler observes the aggregated system state
\begin{equation}
o_t =
\big[
p_{i,t},\;
c_{i,t},\;
q_{i,t}
\big]_{i=1}^C
\;\oplus\;
\big[
\theta_{d,t},\;
\theta^{\mathrm{amb}}_{d,t},\;
\psi_{d,t}
\big]_{d=1}^D,
\end{equation}
yielding an observation dimension of $3C + 3D$.

\subsubsection{Action Space}

The action at time $t$ is given by $a_t = (a^{\mathrm{job}}_t,
a^{\mathrm{cool}}_t)$, consisting of job assignment and cooling control
decisions. For each arriving job $j \in \mathcal{J}_t$, the scheduler selects an assignment
$a^j_t \in \{1,\dots,C\}$, where $a^j_t = i$ assigns the job to cluster $i$, and
$a^j_t = 0$ defers the job to a future time step.
Cooling actions specify datacenter-level temperature setpoints
\begin{equation}
a^{\mathrm{cool}}_t =
\big[
\theta^{\mathrm{target}}_{1,t}, \dots, \theta^{\mathrm{target}}_{D,t}
\big]
\in [\theta_{\min}, \theta_{\max}]^D ,
\end{equation}
which are tracked by local PID controllers that determine the resulting cooling
power $\Phi^{\mathrm{cool}}_{d,t}$. \emph{In our evaluation, dynamic cooling setpoints are optimized only by
MPC-based controllers. Heuristic baselines operate with fixed,
datacenter-specific setpoints and do not actively control cooling.}

\subsection{System Dynamics}

This subsection describes the physical processes governing state transitions,
including workload arrivals, job execution, thermal evolution, cooling control,
throttling, ambient conditions, power consumption, and electricity pricing.

\subsubsection{Time Model and System Indexing}

Each cluster $i$ is characterized by a heat generation coefficient $\alpha_i$
(W per unit of compute), a compute power coefficient $\phi_i$ (W per unit of
compute), and a maximum compute capacity $c^{\max}_i$. Clusters within the same
datacenter share thermal and electrical infrastructure; thermal dynamics and
cooling control are modeled at the datacenter level.

\subsubsection{Workload Model}

At each time step, the environment generates a batch of jobs
$\lambda_t = \{(r_j, d_j, v_j, \tau_j)\}_{j \in \mathcal{J}_t}$, where $r_j$ denotes
required resources, $d_j$ is execution duration, $v_j$ is priority, and
$\tau_j \in \{\text{CPU}, \text{GPU}\}$ specifies hardware affinity. Jobs not
scheduled immediately remain queued. Workload arrivals are instantiated from the Alibaba cluster trace
dataset~\cite{alibaba2018}, providing realistic arrival patterns, execution
durations, resource demands, and hardware constraints observed in production
systems.

\subsubsection{Job Execution Model}
\label{sec:job_execution}

Jobs execute over multiple time steps, with remaining duration evolving as $d^{\mathrm{rem}}_{j,t+1} = d^{\mathrm{rem}}_{j,t} - 1$,
until completion. Cluster utilization is given by
$
u_{i,t} = \sum_{j \in \mathcal{A}_{i,t}} r_j$,
 where $\mathcal{A}_{i,t}$ denotes the set of active jobs assigned to cluster $i$,
subject to the capacity constraint $u_{i,t} \le c_{i,t}$. Completed jobs with
$d^{\mathrm{rem}}_{j,t+1} = 0$ are removed. The queue length $q_{i,t}$ tracks the
number of waiting (unscheduled) jobs associated with cluster $i$.

\subsubsection{Thermal Dynamics}

Datacenter temperature evolves according to a lumped thermal RC model~\cite{pathak2019bayesian}:
\begin{equation}
\theta_{d,t+1}
=
\theta_{d,t}
+
\underbrace{\tfrac{\Delta t}{C_d}
\sum_{i \in \mathcal{C}_d} \alpha_i u_{i,t}}_{\text{Compute Heat}}
\!-\!
\underbrace{\tfrac{\Delta t}{C_d R_d}
(\theta_{d,t} - T^{\mathrm{amb}}_{d,t})}_{\text{Passive Dissipation}}
\!-\!
\underbrace{\tfrac{\Delta t}{C_d}
\Phi^{\mathrm{cool}}_{d,t}}_{\text{Active Cooling}} .
\end{equation}

where $\mathcal{C}_d$ denotes the set of clusters hosted in datacenter $d$, and
$R_d$ and $C_d$ are thermal resistance (°C/W) and capacitance (J/°C),
respectively. This formulation captures thermal inertia linking present
scheduling decisions to future thermal constraints.

\subsubsection{PID Cooling Control}

Cooling power is determined by a PID controller:
\begin{equation}
\Phi^{\mathrm{cool}}_{d,t}
=
\min\!\Big(
\underbrace{\Phi^{\max}_d}_{\text{Cooling Limit}},\;
\underbrace{K_p e_t}_{\text{Proportional}}
\!+\!
\underbrace{K_i \!\sum_{\tau=0}^t e_\tau \Delta t}_{\text{Integral}}
\!+\!
\underbrace{K_d \tfrac{e_t - e_{t-1}}{\Delta t}}_{\text{Derivative}}
\Big),
\end{equation}

where $e_t = \max(0, \theta_{d,t} - T^{\mathrm{target}}_{d,t})$ is the temperature
tracking error, and $K_p$, $K_i$, and $K_d$ are controller gains.

\subsubsection{Thermal Throttling via Soft Constraints}

Sustained high temperatures trigger throttling that degrades effective compute
capacity. For each datacenter $d$, we define thresholds
$\theta_d^{\text{soft}}$ (throttling onset) and $\theta_d^{\max}$ (hard limit).
Effective cluster capacity is
\begin{equation}
c^{\mathrm{eff}}_{i,t} = c^{\max}_i \, g(\theta_{d,t}),
\end{equation}
where
\begin{equation}
c^{\mathrm{eff}}_{i,t}
=
c^{\max}_i\,
\max\!\left(
g_{\min},\;
\min\!\left(
1,\;
1-(1-g_{\min})
\frac{\theta_{d,t}-\theta_d^{\mathrm{soft}}}
{\theta_d^{\max}-\theta_d^{\mathrm{soft}}}
\right)
\right).
\end{equation}

\subsubsection{Ambient Temperature}

Ambient temperature follows a diurnal process:
\begin{equation}
\theta^{\mathrm{amb}}_{d,t} =
\theta_{\mathrm{base},d}
+ A_d \sin\!\left( \frac{2\pi t}{\theta} \right)
+ \epsilon_t,
\quad
\epsilon_t \sim \mathcal{N}(0,\sigma^2),
\end{equation}
capturing climate-dependent outdoor variability. When available, real
temperature traces can be loaded in place of this model.

\subsubsection{Power Dynamics}

Available electrical power capacity evolves as
\begin{equation}
p_{i,t+1}
=
\underbrace{p_{i,t}}_{\text{Available Power}}
\!-\!
\underbrace{\phi_i u_{i,t}}_{\text{Compute Power}}
\!-\!
\underbrace{\kappa_i \Phi^{\mathrm{cool}}_{d(i),t}}_{\text{Cooling Power}}
\!+\!
\underbrace{w_{i,t}}_{\text{Grid Inflow}} .
\end{equation}

where $\kappa_i$ denotes the cluster-level cooling power coupling coefficient,
$d(i)$ is the datacenter hosting cluster $i$, and $w_{i,t}$ represents grid power
inflow. The environment enforces $p_{i,t} \ge 0$ via admission control.

\subsubsection{Electricity Pricing}

Electricity prices vary by location and time. Each datacenter faces time-varying
prices, with $\psi_{d,t} = \psi_d^{\text{peak}}$ for
$t \in \mathcal{T}_{\text{peak}}$ and $\psi_{d,t} = \psi_d^{\text{off-peak}}$
otherwise. Total operational cost is
\begin{equation}
\text{Cost}_t
=
\sum_{d=1}^D
\Big(
\underbrace{\psi_{d,t}\!\sum_{i \in \mathcal{C}_d} \phi_i u_{i,t}}_{\text{Compute Cost}}
\!+\!
\underbrace{\psi_{d,t}\Phi^{\mathrm{cool}}_{d,t}}_{\text{Cooling Cost}}
\Big)
\Delta t .
\end{equation}

\subsubsection{Cluster Dynamics and Job Progression}

Cluster state evolves through utilization updates, power consumption, and job
completion. Completed jobs are removed from execution, and queue lengths are
updated accordingly.

%% file: Allocation.tex
\section{Scheduling Policies for Evaluation}
\label{sec:allocation}

We evaluate a representative set of scheduling policies spanning random,
heuristic, and model-based control approaches to quantify trade-offs among
throughput, queueing latency, thermal safety, and energy efficiency. All
policies operate within the same \textit{DataCenterGym} interface and are
subject to identical feasibility and safety constraints. At each decision
epoch $t$, the scheduler observes the system observation $o_t$ and selects a
job allocation action $a_t \in \mathcal{A}(o_t)$.

\paragraph{Notation}
Jobs are indexed by $j$, clusters by $i$, data centers by $d$, and time by $t$.
Job $j$ requires $r_j$ units of compute of type
$\tau(j)\in\{\mathrm{CPU},\mathrm{GPU}\}$. Cluster $i$ has effective compute
capacity $c_{i,t}$ and utilization $u_{i,t}$, where
$c_{i,t} = c_i^{\max} g(\theta_{d(i),t})$ accounts for thermal throttling.
$\mathcal{C}_d$ denotes the set of clusters hosted in data center $d$, and
$\mathcal{C}_d^\tau \subseteq \mathcal{C}_d$ the subset supporting hardware type
$\tau$. The set $\mathcal{F}(j,o_t)$ denotes clusters feasible for job $j$ under
observation $o_t$, accounting for capacity, hardware affinity, and safety
constraints.

\paragraph{Thermal and power parameters}
$\theta_{d,t}$ denotes data center temperature,
$\theta^{\mathrm{target}}_{d,t}$ the cooling setpoint, 
$\alpha_i$ the heat-generation coefficient, and
$\phi_i$ the compute power coefficient.
The function $g(\theta)\in[0,1]$ denotes the monotone thermal throttling function
defined in Section~\ref{sec:environment}.

\subsection{Random Baseline}

The random baseline assigns each arriving job uniformly at random among feasible
clusters:
\begin{equation}
\mathbb{P}(a^j_t = i \mid o_t) =
\begin{cases}
\frac{1}{|\mathcal{F}(j,o_t)|}, & i \in \mathcal{F}(j,o_t), \\
0, & \text{otherwise}.
\end{cases}
\end{equation}
This policy ignores physical system state and provides a lower-bound baseline.

\subsection{Greedy Capacity-Based Policy}

The greedy policy assigns each job to the feasible cluster with the lowest
normalized utilization:
\begin{equation}
i^\star =
\arg\min_{i \in \mathcal{F}(j,o_t)}
\frac{u_{i,t}}{c_{i,t}}
\quad
\text{s.t.}
\quad
c_{i,t} - u_{i,t} \ge r_j .
\end{equation}
This promotes load balancing and throughput but does not account for thermal
dynamics or energy impact.

\subsection{Thermal-Aware Heuristic}

The thermal-aware heuristic routes jobs to avoid instantaneous thermal
hotspots. For assigning job $j$ to cluster $i$, the estimated post-assignment
data center temperature proxy is
\begin{equation}
\tilde{\theta}_{d(i),t}(j)
=
\theta_{d(i),t}
+
\alpha_i r_j .
\end{equation}
Jobs are assigned to the feasible cluster minimizing
$\tilde{\theta}_{d(i),t}(j)$. This reduces immediate thermal stress but remains
myopic to thermal inertia and future workload evolution.

\subsection{Power and Cooling-Aware Heuristic}

The power--cooling heuristic minimizes instantaneous energy impact. The marginal
incremental power draw of assigning job $j$ to cluster $i$ is approximated by
\begin{equation}
\Delta P_{i,t}(j)
=
\phi_i r_j
+
\omega\,\hat{\Phi}^{\mathrm{cool}}_{d(i),t}(j),
\end{equation}
where $\omega>0$ weights cooling relative to compute power and
$\hat{\Phi}^{\mathrm{cool}}_{d,t}(j)$ estimates incremental cooling power:
\begin{equation}
\hat{\Phi}^{\mathrm{cool}}_{d,t}(j)
\approx
\gamma\Bigl(
\underbrace{\theta_{d,t}-\theta^{\mathrm{target}}_{d,t}}_{\text{Thermal gap}}
+
\underbrace{R_d \alpha_i r_j}_{\text{Heat load}}
\Bigr).
\end{equation}
with gain $\gamma>0$. Jobs are assigned to the feasible cluster minimizing
$\Delta P_{i,t}(j)$. This policy captures instantaneous efficiency but does not
plan over time.

\subsection{Safety-Constrained Model Predictive Control (SC-MPC)}
\label{sec:scmpc}
Model Predictive Control (MPC) solves a finite-horizon constrained optimization problem and replans at each step using new observations. Safety-constrained MPC (SC-MPC)~\cite{hewing2020learning} extends this framework by explicitly enforcing safety-critical limits via hard constraints and penalized soft constraint violations. Formally, following the notation of~\cite{hewing2020learning}, consider discrete-time dynamics
$x(k+1)=f(x(k),u(k);\eta(k))$ with exogenous inputs $\eta(k)$. At time $k$, SC-MPC solves
\begin{align} \min_{\mathbf{U},\boldsymbol{\xi}}\quad & \ell_f(x_{N|k}) + \sum_{i=0}^{N-1} \ell(x_{i|k},u_{i|k},\xi_{i|k}) \\ \text{s.t.}\quad & x_{0|k}=x(k), \\ & x_{i+1|k}=f(x_{i|k},u_{i|k};\hat{\eta}_{i|k}), \\ & u_{i|k}\in\mathcal{U}_{\mathrm{hard}}, \\ & x_{i|k}\in\mathcal{X}_{\mathrm{hard}}, \\ & x_{i|k}\in\mathcal{X}_{\mathrm{soft}}(\xi_{i|k}), \quad \xi_{i|k}\ge 0 . \end{align}
where hard constraints enforce non-negotiable safety limits and soft constraints allow controlled violations through slack variables.

\paragraph{Instantiation in DataCenterGym}

We set $x(k)\equiv o_k$ and $u(k)\equiv a_k$. Predicted states evolve as
\begin{equation}
\hat{o}_{k+i+1}
=
f(\hat{o}_{k+i},a_{k+i};\hat{\eta}_{k+i}),
\end{equation}
where nominal exogenous inputs $\hat{\eta}$ (ambient temperature and electricity
price) effect the system evolution. Hard constraints enforce thermal and
capacity safety:
\begin{align}
\hat{\theta}_{d,k+i} &\le \theta_d^{\max}, \\
0 \le \hat{u}_{i,k+i} &\le c_i^{\max} g(\hat{\theta}_{d(i),k+i}), \\
\theta^{\mathrm{target}}_{d,k+i} &\in [\theta_{\min}, \theta_{\max}].
\end{align}
\subsection{Hierarchical Joint Scheduling and Thermal Control MPC} \label{sec:hierarchical_mpc} We formulate data center operation as a joint scheduling and thermal control problem with a hybrid (discrete--continuous) action space. To achieve tractable optimization at scale, we adopt a Hierarchical MPC (H-MPC) architecture that decomposes control into  (i) a
data center-level supervisory MPC over horizon $H_1$ and (ii) per-datacenter
cluster-level scheduling MPCs over horizon $H_2$. Different horizons reflect slow thermal dynamics versus fast workload dynamics, with $H_2 \le H_1$ ensuring consistency with long-term thermal planning while remaining computationally tractable.

\subsubsection{Composite Action Space}
At each decision epoch $t$, the controller outputs a composite action
$a_t = \bigl(a^{\mathrm{sched}}_t,\; a^{\mathrm{therm}}_t\bigr)$, where
$a^{\mathrm{sched}}_t \in \mathcal{A}_{\mathrm{sched}}$ and
$a^{\mathrm{therm}}_t = \{\theta^{\mathrm{target}}_{d,t}\}_{d=1}^D \in [\theta_{\min}, \theta_{\max}]^D$.

\subsubsection{Stage 1 -- Data Center-Level Supervisory MPC} This stage optimizes admission decisions and thermal setpoints over horizon $H_1$ using aggregate workload variables. Let $\rho_{\tau,k}\in[0,1]$ denote the admission fraction for job type $\tau$, and let $n_{\tau,k}$ be the number of arrivals. The optimization problem tries to minimize total Energy, Queue Length, Thermal Difference from Target setpoints and admission (with soft thermal constraints)   by

\begin{align}
& \min_{\{\rho_{\tau,k},\,\theta^{\mathrm{target}}_{d,k},\,\xi_{d,k}\}}   
\sum_{k=0}^{H_1-1}\Bigl(
\underbrace{\lambda_E E_k}_{\text{Energy}}
+\underbrace{\lambda_Q Q_k}_{\text{Queue/Backlog}}
+\underbrace{\lambda_T \|\theta_{d,k}-\theta^{\mathrm{ref}}\|^2}_{\text{Temperature Deviation}}
\Bigr)\nonumber\\
& +\sum_{k=0}^{H_1-1}\Bigl(
\underbrace{\lambda_R \sum_{\tau} n_{\tau,k}(1-\rho_{\tau,k})}_{\text{Rejection / Low-admission Penalty}}
+\underbrace{\lambda_\xi \sum_d \xi_{d,k}}_{\text{Slack Penalty}}
\Bigr)
\label{eq:stage1_obj}
\end{align}
subject  to box constraints $\rho_{\tau,k}\in[0,1]$, $\theta^{\mathrm{target}}_{d,k}\in[\theta_{\min},\theta_{\max}]$,
and soft thermal limits $\theta_{d,k}\le \theta_d^{\max}+\xi_{d,k}$ with $\xi_{d,k}\ge 0$, we enforce
admission feasibility:

\begin{equation} 
\underbrace{\rho_{\tau,k}n_{\tau,k}}_{\text{Admitted}}
\le
\underbrace{\sum_{i\in\mathcal{C}_d^\tau}\Bigl\lfloor \frac{c_{i,k}-u_{i,k}}{\bar r^\tau} \Bigr\rfloor}_{\text{Max Feasible in DC $d$}},
\ \quad \forall d,\tau,k.
\end{equation}

\subsubsection{Stage 2 -- Cluster-Level Scheduling MPC}
Given fixed setpoints and admission quotas from Stage~1, Stage~2 allocates jobs
to individual clusters over horizon $H_2$. Let $x^{(k)}_{\tau,i}$ denote the
number of type-$\tau$ jobs assigned to cluster $i$ at time $k$. We minimize
(queueing, energy, rejection) by

\begin{align}
\min_{\{x^{(k)}_{\tau,i}\}} \quad
& \sum_{k=0}^{H_2-1}\Bigl(
\underbrace{\lambda_Q \sum_i q_{i,k}}_{\text{Queue/Backlog}}
+
\underbrace{\lambda_E E_k}_{\text{Energy}}
\Bigr)
+
\underbrace{\lambda_{\mathrm{rej}} \sum_{k,\tau} r_{\tau,k}}_{\text{Job Rejection}}
\label{eq:stage2_obj}
\end{align}
subject to
\[
\underbrace{\sum_{i\in\mathcal{C}_d^\tau} x^{(k)}_{\tau,i}}_{ \text{Routed Jobs}}
\!=\!
\underbrace{a^{(k)}_{\tau,d}}_{ \text{Quota}},
\quad
\underbrace{x^{(k)}_{\tau,i}\bar r^\tau}_{  \text{Demand}}
\!\le\!
\underbrace{c_{i,k}-u_{i,k}}_{ \text{Headroom}},
\quad
x^{(k)}_{\tau,i}\in\mathbb{R}_+ .
\]

\subsubsection{Computational Complexity}
\label{sec:time_complexity}

A centralized SC-MPC formulation introduces binary job--cluster assignment
variables over horizon $H$ = $\max(H_1,H_2)$, leading to exponential worst-case
complexity $O(2^{CJH})$, where $C$ denotes the number of clusters and $J$ the
number of jobs per time step. Relaxing these integrality constraints yields a
polynomial-time optimization with complexity $O((CJH)^3)$, which remains
computationally prohibitive at scale. In contrast, H-MPC exploits hierarchical
and geo-distributed structure by decomposing control into a low-dimensional
data center--level supervisory MPC and $D$ parallel per--data center scheduling
subproblems, yielding a per-epoch complexity
$O(D^3H^3) + D\cdot O\!\left((CJH/D^2)^3\right)$,
which is strictly smaller than $O((CJH)^3)$ for any $D>1$ and enables scalable
optimization in large systems.

%% file: Experimental_Setup.tex
\section{Experimental Setup}
\label{sec:experimental_setup}

\begin{table*}[th]
\captionsetup{font=scriptsize}
\centering
\scriptsize
\caption{Experimental configuration (20 clusters across 4 datacenters).
PID gains: $K_p$=4000--7000, $K_i$=80--150, $K_d$=800--1500.
Throttling: $\theta_{\text{soft}}$=32$^\circ$C, $\theta_{\max}$=35$^\circ$C.
Nominal regime: 200 jobs/step, 40/60 CPU/GPU.}
\label{tab:experimental_config}
\begin{tabular}{lcccccc}
\hline
\textbf{DC} &
\textbf{Clusters} &
\textbf{Capacity (CU)} &
\textbf{$\theta_{\text{amb}}$} &
\textbf{\$/kWh} &
\textbf{$R/C$} &
\textbf{Cooling} \\
\hline
Seattle &
3CPU/2GPU &
252K (157C,150G) &
10$\pm$5 &
0.08/0.06 &
0.003/700M &
0.68MW, $g_{\min}$=0.2 \\
&
$\alpha$: CPU[0.3--0.7], CPU[4.0--5.0] &
&
&
&
$T_{1,t}$=23$^\circ$C
& \\

Phoenix &
2CPU/CPU &
235K (65CPU,170G) &
38$\pm$12 &
0.22/0.14 &
0.004/600M &
1.22MW, $g_{\min}$=0.7 \\
&
$\alpha$: CPU[0.6--0.8], CPU[6.5--8.0] &
&
&
&
$T_{2,t}$=25$^\circ$C
& \\

Chicago &
3CPU/2GPU &
204K (144CPU,60GPU) &
16$\pm$10 &
0.13/0.09 &
0.005/550M &
0.30MW, $g_{\min}$=0.4 \\
&
$\alpha$: CPU[0.4--0.6], GPU[3.5--4.5] &
&
&
&
$T_{3,t}$=24$^\circ$C
& \\

Dallas &
2CPU/3GPU &
370K (90CPU,280GPU) &
30$\pm$11 &
0.19/0.11 &
0.002/520M &
1.97MW, $g_{\min}$=0.3 \\
&
$\alpha$: CPU[0.5--0.7], GPU[6.0--9.0] &
&
&
&
$T_{4,t}$=24$^\circ$C
& \\
\hline
\end{tabular}
\end{table*}
We evaluate scheduling policies using Monte Carlo simulation on a
 geo-distributed datacenter topology with heterogeneous
hardware. Unless otherwise stated, all experiments operate in the nominal
regime targeting 60--70\% utilization.

\subsection{Simulation Environment}

All experiments use the \textit{DataCenterGym} environment defined in
Section~\ref{sec:environment}, including thermal dynamics, throttling behavior,
and energy accounting. Here we summarize only the parameters relevant to the
experimental configuration.

\textbf{Thermal Dynamics and Throttling Parameters:}
Datacenters follow the lumped RC thermal model and throttling mechanism described
in Section~\ref{sec:environment}. The temperature state $\theta$ is modeled as a
single scalar per datacenter, serving as a control-level thermal proxy rather
than a physical CPU junction temperature. Accordingly, soft and hard thermal
limits are defined at $\theta_{\mathrm{soft}}=32^\circ$C and
$\theta_{\max}=35^\circ$C, respectively, to induce throttling behavior within
this abstracted model rather than to reflect realistic hardware operating
thresholds (typically $\sim$80$^\circ$C), which are outside the scope of the
lumped building-level RC approximation. Minimum throttling factors $g_{\min}$
vary across datacenters in the range $[0.2,\,0.7]$. Cooling setpoints are
datacenter-specific and fixed in the range 23--25$^\circ$C unless explicitly
optimized by the controller.

\textbf{Energy and cost parameters:}
Energy consumption is computed from compute and cooling power as defined in
Section~\ref{sec:environment}, using a 5-minute timestep ($\Delta t=300$s).
Time-of-use electricity prices vary by location between \$0.06 and \$0.22 per
kWh.

\textbf{Job execution model:}
Jobs follow the capacity-constrained execution model described in
Section~\ref{sec:environment}, with FIFO scheduling and backfilling. Energy
efficiency is reported as total energy divided by completed jobs.

\textbf{Job Completion Tracking:} Each job with resource demand $r$ (CU), duration $d$ (timesteps), and affinity $\tau \in \{\text{CPU}, \text{GPU}\}$ executes when capacity $\geq r$, decrementing $d$ each timestep until completion. Capacity-constrained processing with backfilling: jobs process in FIFO order up to available capacity; if a job doesn't fit, smaller jobs behind it can still execute. Completed jobs ($N_{\text{completed}}$) enable energy efficiency calculation: $E/J = E_{\text{total}} / N_{\text{completed}}$.

\subsection{System Configuration}

Experiments use across 20 heterogeneous clusters ($C$)   and 4  geographically
distributed data centers ($D$). Hardware characteristics, thermal parameters, and
capacity distributions are summarized in
Table~\ref{tab:experimental_config}. Parameters were obtained via a structured
offline calibration process combining physical feasibility constraints,
documented operating ranges, and iterative closed-loop simulation. Automated
tuning workflows and sensitivity analysis were used to efficiently explore the
high-dimensional configuration space and identify stable, realistic operating
points. All parameters were validated under closed-loop control, and no
additional configuration assumptions are introduced beyond those reported in
the table.

\subsection{Workload Trace}

Workloads are derived from the Alibaba 2018 cluster trace~\cite{alibaba2018},
which contains 12.2M jobs over eight days. A contiguous 24-hour slice is
extracted and mapped to 5-minute timesteps (288 per episode) to avoid startup
artifacts.

To target nominal utilization, arrivals are capped at 200 jobs per timestep. Alibaba CPU and memory demands are normalized to compute units (CU)
and scaled to cluster capacities. Because GPU annotations are absent in the
trace, a 40\% CPU / 60\% GPU affinity split is synthesized based on reported
production GPU adoption rates~\cite{barroso2019datacenter}.

\begin{figure*}[t]
\captionsetup{font=scriptsize}
\centering
\begin{minipage}[b]{0.32\linewidth}
\centering
\includegraphics[width=\linewidth]{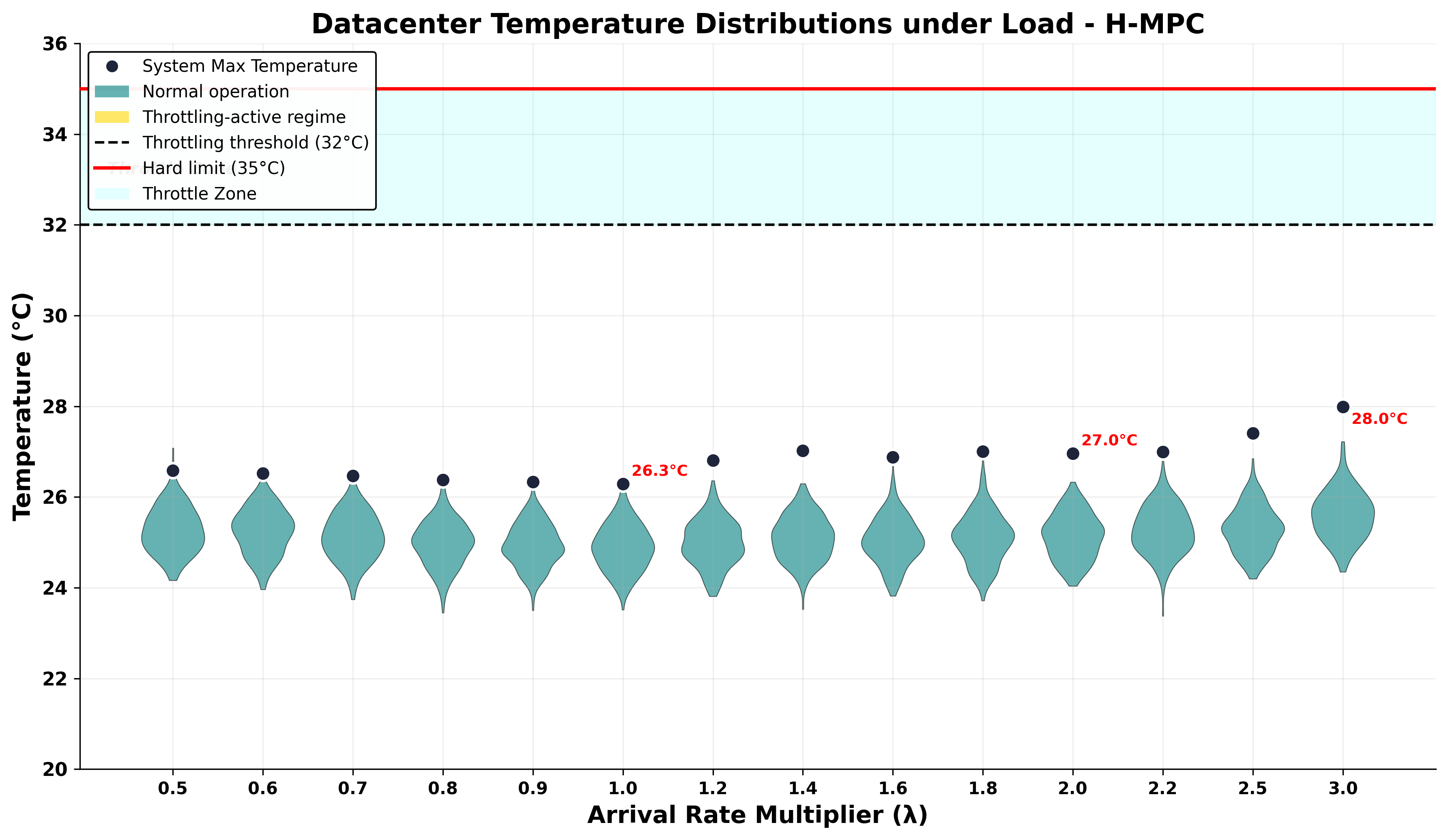}
\par\scriptsize (a) H-MPC
\end{minipage}
\begin{minipage}[b]{0.32\linewidth}
\centering
\includegraphics[width=\linewidth]{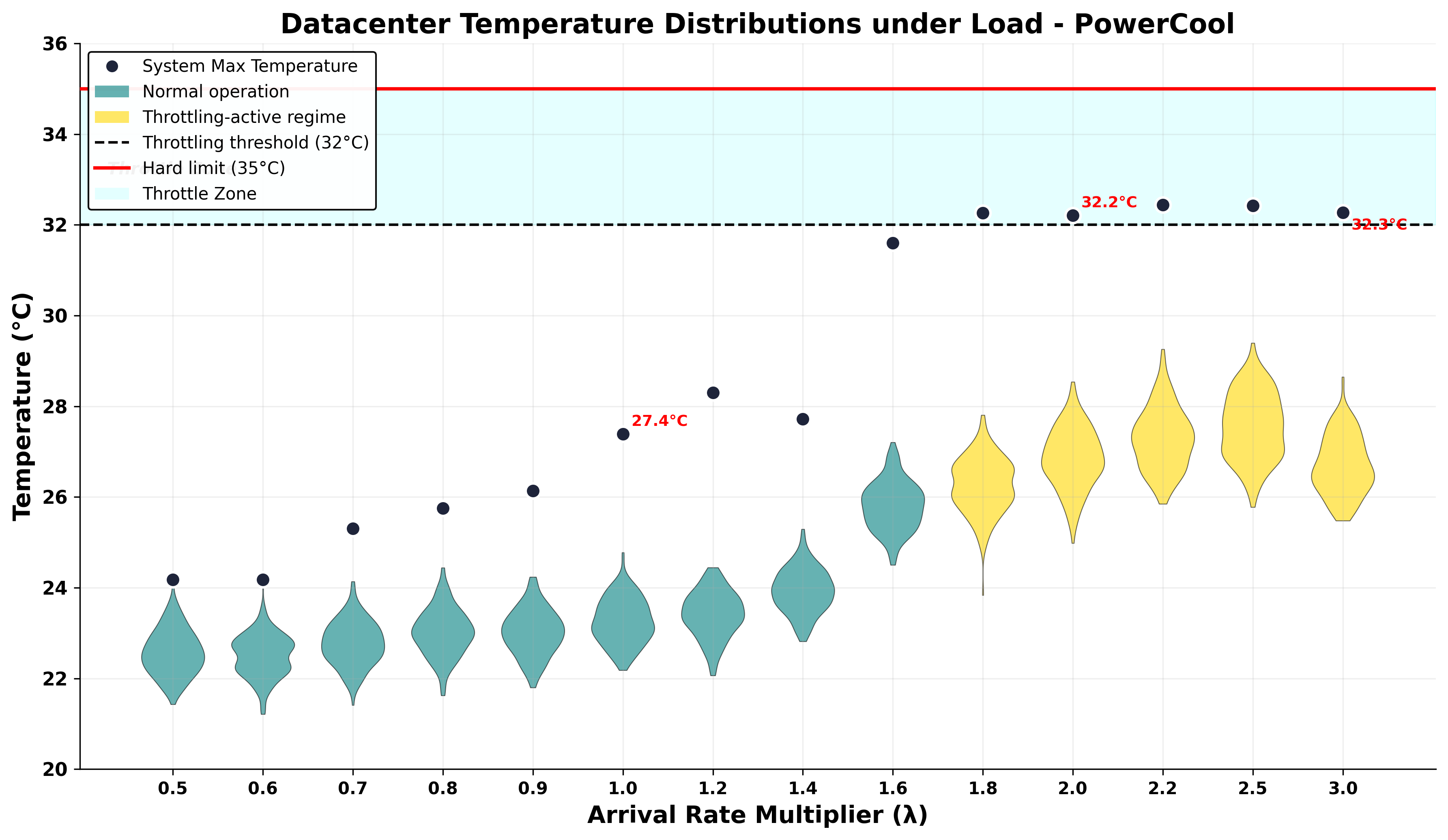}
\par\scriptsize (b) PowerCool
\end{minipage}
\begin{minipage}[b]{0.32\linewidth}
\centering
\includegraphics[width=\linewidth]{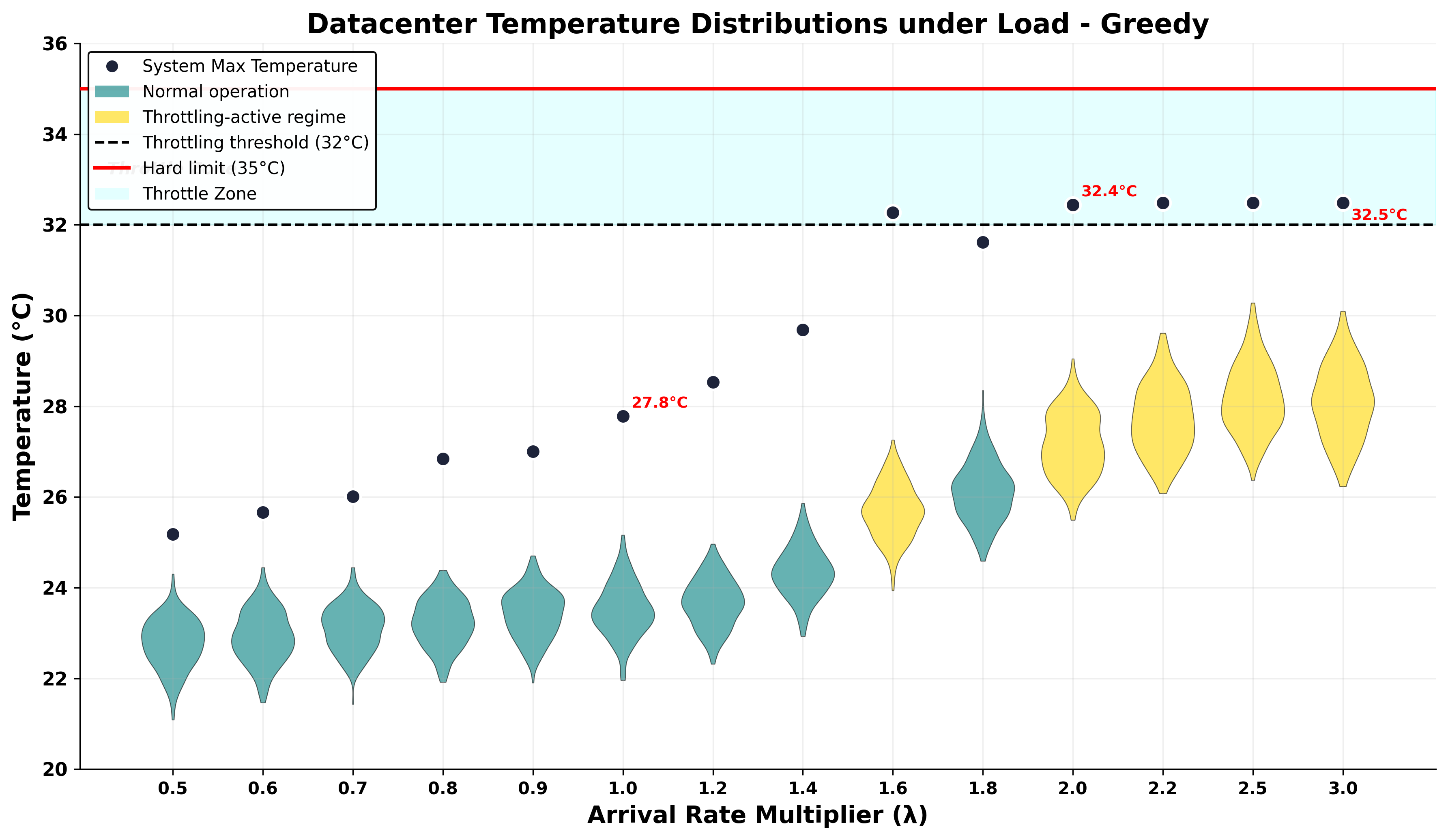}
\par\scriptsize (c) Greedy
\end{minipage}
\caption{Thermal response under increasing workload. H-MPC actively tracks
temperature setpoints, maintaining tightly bounded distributions and preserving
thermal headroom. PowerCool moderates but does not prevent thermal escalation.
Greedy drives the system into the throttling-active regime beyond
$\lambda \approx 1.6\times$. System Temperature indicates temperature around the Cluster which employ the throttling mechanism.}
\label{fig:rq2_thermal_under_load}
\end{figure*}
\subsection{Evaluation Protocol}

\textbf{Episodes:}
Each policy is evaluated over 24-hour episodes (288 timesteps). The system
reaches thermal equilibrium within the first hour; therefore, no warm-up period
is discarded.

\textbf{Statistical Methodology:}
Results are obtained via Monte Carlo evaluation with five independent random
seeds. For each seed, workload arrivals and ambient temperature trajectories
are held fixed across policies. Deterministic policies use seeded tie-breaking.
Metrics are reported as mean $\pm$ standard deviation.

\textbf{Evaluation Metrics:}
Table~\ref{tab:metrics} summarizes the performance metrics spanning QoS,
thermal safety, and energy efficiency.

\begin{table}[h]
\centering
\captionsetup{font=scriptsize}
\scriptsize
\caption{Evaluation metrics}
\label{tab:metrics} 
\begin{tabular}{llp{3.5cm}}
\hline
\textbf{Dimension} & \textbf{Metric} & \textbf{Definition} \\
\hline
\multirow{3}{*}{QoS}
& Utilization ($U_{\text{mean}}$) & Mean CPU/GPU utilization (\%) \\
& Queue Length ($Q_{\text{mean}}$) & Mean queued jobs per cluster \\
\hline
\multirow{2}{*}{Thermal}
& Temperature ($\theta_{\text{mean}}, \theta_{\text{max}}$)
& Mean / max datacenter temperature (°C) \\
& Throttle Time (\%) & \% timesteps with $\theta_{d,t}>\theta_{\text{soft}}$ \\
\hline
\multirow{3}{*}{Energy}
& Total Energy & $E_{\text{total}}$ (kWh) \\
& Energy per Job & $E_{\text{total}}/N_{\text{completed}}$ (kWh) \\
& Cost & \$USD per 24-hour episode \\
\hline
\end{tabular}
\end{table}

%% file: Results.tex
\section{Experimental Evaluation}
\label{sec:results}

We present two representative research questions that DataCenterGym can be used to study.
Unless stated otherwise, all experiments assume the four-datacenter configuration described in Section~\ref{sec:experimental_setup} and employ the candidate scheduling policies introduced in Section~\ref{sec:allocation}.

\textbf{RQ1:} How do scheduling policies compare in the nominal operating regime?  

\textbf{RQ2:} How does system behavior change as job arrival rates increase?

\subsection{RQ1: Nominal Operating Regime}
\label{sec:rq1}

 The nominal operating regime establishes a stable reference environment for calibrating system parameters. All policies are evaluated under identical workloads and dynamics. Non-MPC baselines employ fixed cooling setpoints, whereas MPC-based controllers jointly optimize cooling setpoints as part of the control action. However, in RQ1, safety-constrained MPC (SC-MPC) scheduling is computationally intractable (Sec.~\ref{sec:time_complexity}), and the job scheduler does not scale with increasing decision step sizes. Consequently, we restrict SC-MPC to optimizing cooling setpoints $\theta^{\mathrm{target}}_{d,t}$ only, while delegating job placement to a fixed myopic heuristic.

\begin{table}[H]
\captionsetup{font=scriptsize}
\centering
\caption{Policy comparison in the nominal operating regime (200 jobs/step, bundle\_p088)}
\label{tab:nominal_updated}
\footnotesize
\resizebox{\columnwidth}{!}{%
\begin{tabular}{lcccccc}
\hline
\textbf{Metric} & \textbf{Random} & \textbf{Greedy} & \textbf{Thermal} & \textbf{Power-Cool} & \textbf{SC-MPC} & \textbf{H-MPC} \\
\hline
\multicolumn{7}{c}{\textit{Quality of Service}} \\
CPU Util (\%) & 67.3 & 69.3 & 69.4 & 67.3 & 69.3 & 62.6 \\
GPU Util (\%) & 73.9 & 71.8 & 71.9 & \textbf{75.9} & 71.8 & 70.9 \\
CPU Queue & 491 & 338 & 336 & 349 & 338 & \textbf{324}$^\star$ \\
GPU Queue & 568 & 502 & 502 & 525 & 502 & \textbf{449}$^\star$ \\
\hline
\multicolumn{7}{c}{\textit{Thermal Safety}} \\
$\theta_{\text{mean}}$ & 23.4 & 23.5 & 23.5 & 23.4 & \textbf{22.6}$^\star$ & 23.2 \\
$\theta_{\text{max}}$ & 26.4 & 27.2 & 27.5 & 27.1 & \textbf{25.3}$^\star$ & 26.3 \\
Throttle (\%) & 0 & 0 & 0 & 0 & 0 & 0 \\
\hline
\multicolumn{7}{c}{\textit{Energy Efficiency}} \\
kWh/Job & 2.54 & 2.53 & 2.50 & 2.26 & 2.55 & \textbf{2.20}$^\star$ \\
Cost (\$) & 16,408 & 18,242 & 17,840 & 15,109 & 18,313 & \textbf{14,424}$^\star$ \\
\hline
\end{tabular}}
\end{table}

Table~\ref{tab:nominal_updated} contrasts baseline heuristics, SC-MPC, and the
proposed hierarchical MPC (H-MPC) under identical workloads in the nominal
operating regime. As all policies operate far from thermal and capacity limits,
observed differences primarily reflect control structure rather than
stress-induced behavior.

\begin{itemize}

\item \textbf{Queue regulation via target utilization control:}  
H-MPC achieves the lowest CPU and GPU queue backlogs while explicitly regulating
aggregate load to a target utilization range of 60--70\% (averaging $\sim$65\%).
Datacenter-level admission and thermal planning preserve capacity slack,
reducing congestion before cluster-level scheduling.

\item \textbf{Utilization is a control outcome:}  
Greedy and SC-MPC push utilization toward saturation, which does not yield lower
queueing delay. H-MPC instead trades marginal throughput for stable queue
dynamics; utilization emerges from coordinated admission and cooling decisions
rather than being directly optimized.

\item \textbf{Thermal headroom without conservatism:}  
All policies remain thermally safe with no throttling. SC-MPC maintains lower
temperatures via conservative cooling, increasing energy cost, while H-MPC
dynamically adjusts thermal setpoints to operate closer to safe limits.

\item \textbf{Energy efficiency through coordination:}  
H-MPC achieves the lowest total cost and kWh/job among all policies. Unlike
Power-Cool, which relies on load concentration, H-MPC attains these gains
through coordinated admission and cooling control.

\end{itemize}

\noindent\textbf{Overall Insight:}  
\textit{In the nominal regime, H-MPC operates in a distinct control mode that
jointly shapes workload intensity and thermal conditions upstream, yielding
lower queues and competitive energy efficiency without relying on saturation.}

\subsection{RQ2: Workload Intensity Sensitivity}
\label{sec:rq2}

RQ2 examines how the system transitions from nominal to
capacity-constrained operation as workload increases, and whether predictive
control alters this transition. We test with different arrival rates
$\lambda \in \{0.5,\dots,3.0\}$ under Greedy, PowerCool, and H-MPC.

Greedy exposes the plant’s intrinsic saturation behavior. PowerCool adds a
reactive energy bias. H-MPC optimizes a coupled objective over throughput,
thermal deviation, and cooling effort, actively tracking temperature setpoints
while coordinating admission and scheduling. This enables anticipatory
mitigation of thermal pressure before hard limits are approached.

\begin{figure}[t]
\captionsetup{font=scriptsize}
\centering
\includegraphics[width=0.4\textwidth]{figures/rq2_clean_system_saturation.png}
\caption{System saturation under increasing load. Each curve traces operating
points as arrival rate $\lambda$ increases. Greedy (gold) reveals the plant’s
intrinsic saturation behavior, with a sharp knee marking the onset of
congestion-dominated operation. PowerCool follows the same geometry. H-MPC
(black diamonds) biases operation toward the nominal regime (shaded band),
tracking the 60--70\% target without aggressively chasing saturation.}
\label{fig:arrival_frontier}
\end{figure}

\begin{itemize}

\item \textbf{Intrinsic saturation geometry:}  
Greedy reveals a sharp utilization--congestion transition, with a knee near
$\lambda \approx 1.6\times$. PowerCool follows the similar plant frontier and is slightly inwards than Greedy which implies larger queue length as it attempts to reduce.

\item \textbf{Nominal-regime tracking:}  
H-MPC remains near the target operating band across the sweep, delaying entry
into the congestion-dominated region.

\item \textbf{Anticipatory thermal control:}  
H-MPC suppresses temperature excursions before throttling is triggered,
preserving headroom under load. PowerCool remains reactive; Greedy drives the
system into thermal stress.

\end{itemize}

\noindent\textbf{Overall Insight:}  
\textit{Workload scaling exposes an intrinsic utilization--congestion transition:
heuristic policies ride the plant frontier toward saturation and thermal stress.
H-MPC instead couples workload admission with predictive thermal control,
biasing operation toward the nominal regime and preserving thermal headroom as
load increases—effectively expanding the system’s safe operating envelope.}

%% file: Conclusion.tex
\section{Limitations and Future Work}

\textit{DataCenterGym} makes simplifying assumptions to prioritize interpretability and
experimental control, each of which motivates future extensions.

\begin{itemize}
    \item \textbf{Thermal abstraction:}
    While we employ a lumped RC-circuit thermal model in this work, this abstraction cannot capture fine-grained spatial and temporal thermal dynamics.  Modeling realistic building behavior ultimately requires multi-zone thermal representations, which we leave to future work.

    \item \textbf{Network abstraction:}
    Inter-datacenter latency, bandwidth, and data locality are not modeled.

    \item \textbf{Queueing model:}
    Queues are aggregated without job dependencies or resource fragmentation;
    richer service and dependency-aware scheduling is a natural extension.

    \item \textbf{Learning-based control:}
    We assume known dynamics and rely on model-based MPC; integrating
    learning-based or learning-augmented control to address model mismatch and
    non-stationarity is an important direction.

    \item \textbf{Energy supply:}
    Power availability is treated exogenously; future work will integrate grid and
    renewable supply constraints .

  \item \textbf{Experimental scope:}
More   stress testing (e.g., sharp job-arrival surges, sustained overload, and heavy-tailed service times) is left for future work.

\end{itemize}

\section{Conclusion}
We presented \textit{\textbf{DataCenterGym}}, a physically grounded simulation
framework for studying job scheduling in geo-distributed data centers with
coupled compute, thermal, and power dynamics. The simulator explicitly captures
closed-loop interactions between workload placement, heat generation, cooling,
and performance, and trace-driven experiments show that these couplings
materially affect trade-offs among key performance indicators. We also showed
that a hierarchical MPC (H-MPC) design, which separates long-horizon
datacenter-level thermal planning from short-horizon, feasibility-constrained
cluster-level scheduling, enables anticipatory control at scale. 
\textit{DataCenterGym} is intended as a controlled evaluation environment for
multi-objective data center scheduling and control.